\documentclass[12pt]{article}

\usepackage{amssymb}
\def\labelmark{}
\def\void{}

{\ifx\void\labelname\def\junk{\end{displaymath}}
\else\def\junk{\end{eqnarray}}\fi\junk\labelmark\def\labelname{}}

\newcommand{\bra}{\begin{array}}
\newcommand{\era}{\end{array}}
\newcommand{\beq}{\begin{equation}}
\newcommand{\eeq}{\end{equation}}
\newcommand{\bqn}{\begin{eqnarray}}
\newcommand{\eqn}{\end{eqnarray}}

\font\mybb=msbm10  at 12pt
\def\bb#1{\hbox{\mybb#1}}

\font\mybbi=msbm10  at 9pt
\def\bbi#1{\hbox{\mybbi#1}}

\def\BC{\bb C}
\def\_\BC{\bbi C}

\newcommand{\om}{\omega}

\newcommand{\eps}{\epsilon}

\newcommand{\be}{\beta}
\newcommand{\vap}{\varphi}

\newcommand{\te}{\theta}

\newcommand{\pa}{\partial}
\newcommand{\al}{\alpha}

\newcommand{\ov}{\over}
\newcommand{\hb}{\hbar}

\newcommand{\ev}{\equiv}
\newcommand{\lb}{\label}

\newcommand{\PR}[1]{ {\it Phys.~Rev.} {\bf #1}}
\newcommand{\PRL}[1]{ {\it Phys.~Rev.~Lett.} {\bf #1}}

\newcommand{\JMP}[1]{ {\it J. Math.~Phys.} {\bf #1}}

\begin{document}
\begin{titlepage}
\setcounter{page}{1}
\renewcommand{\thefootnote}{\fnsymbol{footnote}}


\vspace{6mm}
\begin{center}

{\large\bf 
A Matrix Model for Bilayered Quantum Hall Systems}

\vspace{8mm}
{\bf Ahmed Jellal
\footnote{E-mail: {\textsf jellal@gursey.gov.tr }}
\footnote{Permanent address: {\textsf Physics Department, Faculty of Sciences, 
Chouaib Doukkali University, Road Ibn M\^aachou P.O. Box 20, 
El Jadida, Morocco }}}
and
{\bf Michael Schreiber }

\vspace{5mm}
{\em  Institut f\"{u}r Physik, Technische 
Universit\"{a}t,\\ D-09107 Chemnitz, Germany }\\

\end{center}

\vspace{5mm}
\begin{abstract}
We develop a matrix model to describe bilayered quantum Hall
fluids for a series of filling factors. Considering two coupling
layers, and starting from a corresponding action, we construct its
vacuum configuration at $\nu=q_iK_{ij}^{-1}q_j$, where $K_{ij}$ is
a $2\times 2$ matrix and $q_i$ is a vector. Our model allows us to
reproduce several well-known wave functions. We show that the wave
function $\Psi_{(m,m,n)}$ constructed years ago by Yoshioka,
MacDonald and Girvin 
for the fractional quantum Hall effect at filling factor 
${2\over m+n}$ and in particular
$\Psi_{(3,3,1)}$ at filling ${1\over 2}$ can be obtained from our vacuum
configuration. The unpolarized Halperin wave function 
and especially that for the fractional quantum Hall
state at filling factor ${2\over 5}$ can also be recovered from
our approach. Generalization to more than $2$ layers is
straightforward.
\end{abstract}
\end{titlepage}

\newpage

\section{Introduction} 

The quantum Hall (QH) effect has bred much beautiful theory. 
Indeed, Laughlin's wave functions~\cite{laughlin}
are good wave functions for describing
the fractional quantum Hall effect (FQHE)\cite{gir}
at filling factor $\nu={1\ov m}$, where $m$ is an odd integer. 
For other filling factors 
several attempts have been suggested to extend Laughlin's theory
by adopting different approaches and assumptions. 
In particular, Halperin~\cite{halperin} proposed
a family of generalized Laughlin wave functions
that could incorporate reversed spins. In fact a candidate for an
unpolarized wave function at filling factor ${2\ov 5}$ was given. 
Subsequently, Yoshioka, MacDonald and Girvin~\cite{girvin1} 
generalized the Laughlin wave functions to those
of the bilayered QH systems and derived that corresponding 
to the $\nu={1\ov 2}$ state. Moreover, other theories
have been elaborated and led to understand the observed
values of  $\nu$, in particular
 $\nu={5\ov 2}$~\cite{read} 
as well as others~\cite{GROUP0}.

The first experimental indications for an unpolarized ground-state spin 
configuration in the FQHE came with the discovery  
of the $\nu={5\ov 2}$ state~\cite{WILLETT}
and later the $\nu={4\ov 3}$ state~\cite{GROUP1}.
More compelling evidence for novel spin phenomena in the FQHE was
subsequently reported~\cite{GROUP2}. On the other hand, it was 
shown experimentally that multi--layer systems also exhibit the 
FQHE~\cite{sarma}. In fact, several filling 
factors have been observed, for instance  
the $\nu={1\ov 2}$  state~\cite{GROUP3} and 
$\nu={9\ov 2}, {11\ov 2},\cdots$~\cite{GROUP22}.

Recently, Susskind~\cite{susskind} proposed a novel method
to investigate the FQHE. He claimed that the non-commutative 
Chern-Simons theory (NCCS) at level $k$ is exactly equivalent to 
Laughlin's theory at the filling factor $\nu_{\rm S}={1\ov k}$. 
He formulated his approach as a matrix theory similar to that describing 
$D0$-branes in string theory. 
However, Susskind's theory is an alternative approach 
to the FQHE which so far has not produced anything new but has just 
recovered the Laughlin approach by adopting a new formalism. 
Nevertheless
it remains a new way of thinking
and is 
worth studying in the hope that it will 
bring new results in the future.

Although the proposed matrix model
seems to reproduce the basic features of the Laughlin QH droplets,
still some problems remain to be solved. Indeed, Susskind's approach is
valid only for infinite matrices and also shows an anomaly for $k=0$. 
To solve these problems, Polychronakos~\cite{polychronakos}
introduced a boundary term to Susskind matrix model.
He proposed a finite matrix model as a regularized version
of the NCCS theory. It allowed him to find
a quantum correction to $\nu_{\rm S}$, where $k$ is shifted to
$k+1$ and the filling factor became $\nu_{\rm PS}={1\ov k+1}$. 
As another consequence, he pointed out that his matrix model 
is equivalent to the Calogero model~\cite{calogero}.

Sometimes later, observing that the Laughlin wave functions 
can be mapped onto many-body wave functions of the harmonic
oscillator, Hellerman and Van Raamsdonk~\cite{hellerman1}
built a complete minimal basis of wave functions of the theory at 
arbitrary level $k$ and rank $M$, see also~\cite{hellerman2}. 
Other investigations about the relation between NCCS and
Laughlin fluids can be found in~\cite{sakita,kobashi}. 
Subsequently, the Susskind model and its regularized version
introduced by Polychronakos was extended to FQH states that are 
not of Laughlin type: a multicomponent Chern-Simons approach was 
introduced~\cite{saidi1} and another proposal based on the 
Haldane hierarchy~\cite{haldane} was developed~\cite{jellal2,jellal3}.

Despite the progress in the study of the FQH fluids in the framework
of NCCS matrix model, several open questions remain which
have not been addressed so far. One of these questions
concerns the wave functions that are not of Laughlin type. 
In fact there are many wave functions, that have been 
constructed years ago, e.g. by
Yoshioka {\it et al.}, Halperin, $\cdots$, 
but cannot be recovered by what is
developed so far.

In what follows we propose a matrix model to investigate
the possibility to obtain two of those wave functions. 
This can be done by extending the Susskind--Polychronakos model 
to deal with the QH fluids at the filling factor~\cite{wen}
\begin{equation}
\label{biff}
\nu=q_iK_{ij}^{-1}q_j
\end{equation}
where $K_{ij}$ is an $N\times N$ matrix and $q_i$ is a vector. 
The basic idea is to consider several Susskind--Polychronakos 
systems, let us say $M$ systems, with an interaction between 
them and suppose that all systems possess the same number of particles. 
In the QHE language, this picture is equivalent
to considering multi-layered systems. Without loss of generality,
we fix $M=2$, but as we will see later our analysis
can directly be extended to the generic case  $M\geq 3$.

We start by writing down an appropriate action
as a sum of two terms, for the free and the interacting part.
Subsequently, we derive the corresponding
Hamiltonian, which of course contains 
an interaction. Using a unitary transformation, 
we show that this Hamiltonian can be transformed to
a diagonalized one. Next, we determine the vacuum 
configuration that allows us to recover two
different states. Indeed,
we show that how
the Yoshioka--MacDonald--Girvin wave functions 
at the filling factor $\nu={2\ov m+n}$ can be obtained
from our model and in particular that describing the FQHE
at $\nu={1\ov 2}$. Moreover, 
the unpolarized Halperin wave functions will be derived and 
especially that corresponding to $\nu={2\ov 5}$ state.

In section 2 we recall briefly the NCCS matrix model description of 
the Laughlin fluid. In section 3, we propose an 
action describing a system of two layers,
we derive the Gauss law constraint 
as well as the equations of motion for the different variables. A quantum 
mechanical analysis will be the subject of next section, where
we develop a Hamiltonian that corresponds to the
system under consideration. Under rotation,
we define a set of matrices of harmonic--oscillator operators to
diagonalize the system. In section 5, we build
the vacuum configuration that satisfies the
constraint. A link with literature
will be discussed in the last section where the two
wave functions mentioned above will be recovered. 
We conclude our paper by putting some questions
to be investigated in forthcoming works.

\section{Chern-Simons matrix model} 

Starting from the matrix formulation of
a two-dimensional system with a large number
of electrons in the presence of a perpendicular magnetic field
$B$, 
Susskind~\cite{susskind} showed that the resulting effective
theory is a non-commutative $U(1)$ Chern--Simons gauge theory
at level $k=B\te$. As a consequence, he found
a relation 
\beq
\rho={1\ov 2\pi \te}
\eeq 
which links the non-commutivity 
parameter $\te$ to the density of electrons $\rho$.
By using the definition of the filling factor
\beq
\nu= {2\pi\rho\ov B}
\eeq
in the system of units $(\hb,e,c)$, it is easily seen that 
the fraction $\nu$ can be written in terms of 
the parameter $\te$ as
\beq
\nu = {1\ov B\te}.
\eeq
This beautiful relation is one of the interesting results obtained
recently by Susskind in dealing with the FQH fluids.

Moreover, by exploring the possibility to develop a consistent
finite matrix model for the description of the FQH droplet,
Polychronakos~\cite{polychronakos} suggested to include a new
field into the Susskind model. The proposed action is given by
\begin{equation}
\label{act}
S = \int dt\; {B \over 2}\; {\rm Tr} 
\left\{ \epsilon^{ab} \left(\dot{X} _a + i [A_0,~ X_a]\right) X_b + 2
\theta A_0 - \omega X_a ^2\right\} 
+ \psi^{\dag}\left(i \dot{\psi} - A_0 \psi\right)
\end{equation}
where $X_a,~a=1,2$ are $N \times N$ matrices and 
$\psi$ is a complex $N$-vector, and $\eps^{12}=-\eps^{21}=1$,
$\eps^{aa}=0$.
The action is invariant under
the gauge group $U(N)$ and the matrix model variables transform
as 
\begin{equation}
X_a \rightarrow U X_a U^{-1},~~~~~~~~~
\psi \rightarrow U \psi.
\label{transf}
\end{equation}
The equation of motion for $A_0$ leads to the 
Gauss law constraint
\begin{equation}
G \equiv -iB \left[X_1, X_2\right] + \psi \psi^{\dag} -B \theta =0.
\label{constraint}
\end{equation}
The trace of this equation gives
\begin{equation}
\psi^{\dag} \psi = N B \theta.
\label{trace}
\end{equation}

Upon quantization the matrix elements of $X_a$ and the components of 
$\psi$ become operators, obeying the commutation relations
\begin{equation}
\begin{array}{l}
\left[ \psi_i, \psi_j^{\dag} \right]  =  \delta _{ij} \nonumber  \\
\left[ (X_1 ) _{ij}, (X_2 ) _{kl} \right]  =  
{i \over B} \delta _{il} ~\delta _{jk}.
\label{CR}
\end{array}
\end{equation}
The Hamiltonian can be obtained from~(\ref{act}) as
\begin{equation}
H = \omega \left( { N^2 \over 2} + \sum A_{nm}^{\dag} A_{mn}\right)
\label{hamiltonian}
\end{equation}
where the $N\times N$ matrix of harmonic-oscillator 
operators is defined by
\begin{equation}
A_{nm} = \sqrt{B \over 2} \left(X_1 + i X_2\right)_{nm}.
\end{equation}
The corresponding wave function is~\cite{hellerman1}
\begin{equation}
| k \rangle = \left[ \epsilon ^{i_1 ... i_N} \psi^{\dag}_{i_1} 
(\psi^{\dag}A^{\dag})_{i_2}\cdots(\psi^{\dag} A^{\dag N-1})_{i_N} 
\right]^k | 0 \rangle
\label{state}
\end{equation}
where the vacuum $| 0 \rangle$ is annihilated 
by $A$'s and $\psi$'s and
$\epsilon$ is the fully 
antisymmetric tensor. This is a physical
state and therefore satisfies the relation
\beq
G | k \rangle =0.
\eeq 
It is similar to the Laughlin
wave function~\cite{laughlin} at 
the filling factor 
\begin{equation}
\nu={1\ov k+1}.
\end{equation}

Subsequently, one of us and others~\cite{jellal2,jellal3} 
generalized the above results to any filling factor 
which can be expressed as
\begin{equation}
\nu _{k_{1} k_{2}}=\frac{1}{k_{1}}+\frac{1}{k_{2}}
\end{equation}
and in particular to level two of the Haldane 
hierarchy~\cite{haldane}
\beq
\nu _{p_{1} p_{2}} = {p_2\over p_1p_2 -1}
\eeq
by setting 
\beq
k_1=p_1, \qquad k_2=p_1(p_1p_2 -1)
\eeq
where $p_1$ is an odd and $p_2$ is an even integer.

\section{Two coupling matrices model}

We consider two systems with a total number
of particles $M_1+M_2$ which  interact with
each other. Such systems can be seen like  
two coupling layers $i$ containing $M_i$ particles. 
The appropriate action to describe the FQH fluids 
of the whole system at filling factor~(\ref{biff}), is given by
\begin{eqnarray}
\lb{TACT}
S= \int dt \sum_j {K_{jj}\ov 2\te}\; {\rm Tr} 
\left\{ \epsilon^{ab} \left(\dot{X}_a^{(j)} + 
i \left[A_0,~ X_a^{(j)}\right]\right) X_b^{(j)} 
+ 2\theta A_0 -\omega_j \left(X_a^{(j)}\right)^2 \right\} 
\nonumber \\
\ \ \ \ 
+  \psi^{(j)\dag}\left(i \dot{\psi}^{(j)} - A_0\psi^{(j)}\right)
+ \int dt\; K_{12}\; \left\{ {\om_{12}\ov\te}\;
{\rm Tr}\left(X_a^{(1)}X_a^{(2)}\right)
+  \psi^{(1)}\psi^{(2)} \right\}
\end{eqnarray}
which involves two copies of the single-layer action~(\ref{act})
forming the free part. It also contains an interacting part, where
the scalar $K_{12}$ is playing the rule of a coupling parameter 
between the layers $1$ and $2$. The ratio ${K_{jj}\ov \te}$ is
basically the magnetic field $B$. 

It is clear that for $K_{12}=0$, 
the total system becomes decoupling. 
Note that as far as the total
action is concerned, the full gauge symmetry is
$U(M_1)\times U(M_2)$. The matrix model variables transform 
under this invariance as
\begin{equation}
X_a^{(i)} \rightarrow U X_a^{(i)} U^{-1},~~~~~~~~~
\psi^{(i)} \rightarrow U \psi^{(i)}.
\label{ctransf}
\end{equation}
Compared to the original matrix model, there is
the potential term 
\beq
\lb{copot}
V= \sum_j {K_{jj}\ov 2\te} \om_j {\rm Tr} \left(X_a^{(j)}\right)^2 - 
{K_{12}\ov\te} \om_{12} {\rm Tr} \left(X_a^{(1)}X_a^{(2)}\right)
\eeq
analogous to the potential of two coupled 
harmonic oscillators~\cite{HAN} in two-dimensional space. 
This provides a Hamiltonian for the theory.
 
The Gauss law constraint can be obtained by evaluating 
the equation of motion for $A_0$. 
In our case it reads
\begin{equation}
\lb{glpo}
{\cal G}\ev -iK_{11}\left[X_{1}^{(1)}, X_{2}^{(1)}\right] -
iK_{22} \left[X_{1}^{(2)}, X_{2}^{(2)}\right]
+ \left( \psi^{(1)} \psi^{(1)\dag} + 
\psi^{(2)} \psi^{(2)\dag} - K_{11} - K_{22}\right)=0
\end{equation}
where its trace gives
\begin{equation}
\psi^{(1)\dag} \psi^{(1)} + \psi^{(2)\dag} \psi^{(2)} = 
M_1K_{11} + M_2K_{22}.
\end{equation}
Other equations of motion can also be calculated.
For the $X$'s we get
\begin{equation}
\begin{array}{l}
K_{11}\eps^{ab} \dot{X}_a^{(1)} + K_{11}\om_1 
{X}_a^{(1)}
 + K_{12}\om_{12} {X}_a^{(2)}=0
\\
K_{22}\eps^{ab} \dot{X}_a^{(2)} + 
K_{22}\om_2 {X}_a^{(2)}
+ K_{12}\om_{12} {X}_a^{(1)}=0
\end{array}
\end{equation}
while for the $\psi$'s we obtain
\begin{equation}
\begin{array}{l}
i\psi^{(1)\dag} +  K_{12} \psi^{(2)}=0
\\ 
i\psi^{(2)\dag} + K_{12} \psi^{(1)}=0.
\end{array}
\end{equation}
Of course the last set of equations shows 
a difference with respect to the decoupled
case. It can be solved by using a
unitary transformation.

\section{Hamiltonian formalism}

Let us now consider the proposed model quantum
mechanically. We proceed by determining the
total Hamiltonian, which describes the system under consideration. 
It can be obtained from the relation
\beq
{\cal H} = {\dot X}{\pa{L}\ov \pa{\dot X}} - L
\eeq
where ${\pa{L}\ov \pa{\dot X}}$ defines the conjugate momentum. 
This leads to a Hamiltonian as the sum of the free and 
the interacting parts as
\beq
\lb{CHAM1}
{\cal H} = \sum_j {K_{jj}\ov 2\te} \om_j {\rm Tr} 
\left(X_a^{(j)}\right)^2 - 
{K_{12}\ov\te} \om_{12} {\rm Tr} \left(X_a^{(1)}X_a^{(2)}\right)
\eeq
which is nothing but the confining potential~(\ref{copot}). This
means that the kinetic energy is negligible compared to $V$.

It is clear that this form of ${\cal H}$ cannot
be diagonalized directly. Nevertheless, ${\cal H}$
can be transformed to another, factorizing
Hamiltonian ${\cal H}'$ . Probably the best way to do this
is to perform a rotation by a mixing angle $\vap$, of the
$X$'s to new matrices
\begin{equation}
\begin{array}{l}
\lb{trans}
Y_a^{(1)} = X_a^{(1)} \cos{\vap\ov 2} - X_a^{(2)} \sin{\vap\ov 2}
\\
Y_a^{(2)} = X_a^{(1)} \sin{\vap\ov 2} + X_a^{(2)} \cos{\vap\ov 2}.
\end{array}
\end{equation}
It can easily be checked that 
this rotation is a unitary transformation.
Inserting~(\ref{trans}) into~(\ref{CHAM1}), 
one can show that ${\cal H}$ transform to
\beq
\lb{CHAM2}
{\cal H}' = \al {\rm Tr} \left(Y_a^{(1)}\right)^2 + 
\be {\rm Tr} \left(Y_a^{(2)}\right)^2
\eeq 
if the rotating angle satisfies the relation
\beq
\tan \vap= {K_{12}\om_{12}\ov K_{11}\om_{1} -K_{22}\om_{2}}.
\eeq 
The parameters $\al$ and $\be$ are given by
\begin{equation}
\begin{array}{l}
\al= {1\ov\te} \left( K_{11}\om_1 \cos^2{\vap\ov 2}
+ K_{22}\om_2 \sin^2{\vap\ov 2} - {1\ov 2} K_{12}\om_{12} 
\sin\vap\right) 
\\
\be =
{1\ov\te} \left( K_{11}\om_1 \sin^2{\vap\ov 2}
+ K_{22}\om_2 \cos^2{\vap\ov 2} + {1\ov 2} K_{12}\om_{12} 
\sin\vap\right). 
\end{array}
\end{equation}

To diagonalize ${\cal H}'$,
we define two couples of creation and annihilation
matrices of harmonic oscillator operators, 
\begin{equation}
\begin{array}{l}
\lb{creal}
C_{nm}^{(1)} = \sqrt{\al \over 2} (Y_1^{(1)} + i Y_2^{(1)})_{nm}
\\
C_{nm}^{(2)} = \sqrt{\be \over 2} (Y_1^{(2)} + i Y_2^{(2)})_{nm}.
\end{array}
\end{equation}
They satisfy the commutation relations 
\begin{equation}
\begin{array}{l}
\lb{heisa}
\left[C_{nm}^{(1)}, C_{n'm'}^{(1)\dag}
\right]=\delta_{nm'}\delta_{n'm}
\\
\left[C_{ij}^{(2)}, C_{i'j'}^{(2)\dag}
\right]=\delta_{ij'}\delta_{i'j},
\end{array}
\end{equation}
while all others commutators vanish. Now  
${\cal H}^{'}$ can be rewritten as  
\begin{equation}
\lb{ham1}
 {\cal H}^{'}={\al\ov 2} 
\left(2 {\cal M}_{1}+ M_{1}^{2}\right)
+ {\be\ov 2} 
\left(2 {\cal M}_{2}+ M_{2}^{2}\right)
\end{equation}
where the number operators 
\begin{equation}
\bra{l}
{\cal M}_{1}= \sum_{n ,m=1}^{M_{1}} 
C_{mn}^{(1)\dag} C_{nm}^{(1)}\\  
{\cal M} _{2 }=
\sum_{i,j=1}^{M_{2}}
C_{ij}^{(2)\dag} C_{ji}^{(2)}
\era
\end{equation}
are counting the $M_1$ and $M_2$ particles. Thus
under the unitary transformation
the system became decoupling.

\section{Ground-state wave functions}

To begin we emphasize a difference
between the ground state of two coupled 
harmonic oscillators in terms of the coordinates $x_i$ and 
that in terms of their mapped representations $y_i$. 
The wave function
\begin{equation}
\label{YWAV}
\psi_0(\vec{y}) \sim 
\exp{\left\{-\al y^{2}_{1} -\be y^{2}_{2} \right\} } 
\end{equation} 
is separable in the variables $y_{1}$ and $y_{2}$.
However, for the variables $x_{1}$ and $x_{2}$, 
the wave function~(\ref{YWAV}) reads 
\beq
\lb{XWAV}
\psi_{0}(\vec{x}) \sim 
\exp\left\{-\al \left(x_{1}\cos{\vap\over 2} -
x_{2} \sin{\vap\over 2}\right)^{2} 
- \be \left(x_{1}\sin{\vap\over 2} +
x_{2}\cos{\vap\over 2}\right)^{2} \right\}.
\eeq

Next, we will see how these ground states
can be extended to the matrix model formalism.
We begin to determine that for the matrices $Y$. 
By transforming the Gauss law constraint
to the variables $Y$, i.e.
\begin{eqnarray}
\lb{gtran}
\left( K_{11} \cos^2{\vap\ov 2}+ K_{22} \sin^2{\vap\ov 2} \right)
\left[Y_{1}^{(1)}, Y_{2}^{(1)}\right] +  \left( K_{11} \sin^2{\vap\ov 2}
+ K_{22} \cos^2{\vap\ov 2}\right) 
\left[Y_{1}^{(2)}, Y_{2}^{(2)}\right] +
\nonumber\\
{1\ov 2}\left(K_{11} - K_{22}\right) \sin{\vap}
\left\{ \left[Y_{1}^{(1)}, Y_{2}^{(2)}\right] + 
\left[Y_{1}^{(2)}, Y_{2}^{(1)}\right] \right\}
= i\te \left( K_{11} + K_{22} - \phi^{(1)} \phi^{(1)\dag} 
- \phi^{(2)} \phi^{(2)\dag} \right)
\end{eqnarray}
where the Polychronakos fields are also rotated
to new fields
\beq
\bra{l}
\lb{nepoly}
\phi^{(1)} = \psi^{(1)} \cos{\vap\ov 2} - \psi^{(2)} \sin{\vap\ov 2}\\
\phi^{(2)} = \psi^{(1)} \sin{\vap\ov 2} + \psi^{(2)} \cos{\vap\ov 2}.
\era
\eeq
For simplicity let us fix $K_{11}=K_{22}=K$, 
then~(\ref{gtran}) becomes
\begin{equation}
\lb{ngtra}
\left[Y_{1}^{(1)}, Y_{2}^{(1)}\right] + 
\left[Y_{1}^{(2)}, Y_{2}^{(2)}\right] 
= 2iK\te \left( 1- {1\ov 2K}\phi^{(1)} \phi^{(1)\dag} 
- {1\ov 2K}\phi^{(2)} \phi^{(2)\dag} \right).
\end{equation}
Now it is clear that the ground state is simply
a tensor product between those states corresponding
to each layer
\begin{eqnarray}
\label{yvacu}
\lefteqn{
| K \rangle= 
\left[ \epsilon ^{i_1...i_{M_1}} \phi^{(1)\dag}_{i_1} 
(\phi^{\dag(1)}C^{(1)\dag})_{i_2} \cdots
(\phi^{(1)\dag} C^{(1)\dag M_1-1})_{i_{M_1}} \right]^{K} }
\nonumber\\
&
~~~~~~\left[ \epsilon ^{j_1...j_{M_2}} \phi^{(2)\dag}_{j_1} 
(\phi^{(2)\dag}C^{(2)\dag})_{j_2} \cdots
(\phi^{(2)\dag} C^{(2)\dag M_2-1})_{j_{M_2}} \right]^{K}
 | 0 \rangle.
\end{eqnarray}

The ground state~(\ref{yvacu}) can be mapped in terms of the operators of 
the matrices $X$ by expressing  the matrices $C$ of harmonic-oscillator 
operators in terms of those corresponding to the matrices $X$. 
Using~(\ref{trans}) one can show that~(\ref{creal}) takes the form  
\begin{equation}
\begin{array}{l}
\lb{acrel}
C_{nm}^{(1)} = \sqrt{\al \over B} \left( A^{(1)}\cos{\vap\ov 2} 
- A^{(2)} \sin{\vap\ov 2}\right)_{nm}\\
C_{nm}^{(2)} = \sqrt{\be \over B} \left( A^{(1)} \sin{\vap\ov 2} 
+ A^{(2)} \cos{\vap\ov 2}\right)_{nm}
\end{array}
\end{equation}
where the operators  
\begin{equation}
\begin{array}{l}
\lb{area}
A_{nm}^{(1)} = \sqrt{B \over 2} \left(X_1^{(1)} + i X_2^{(1)}\right)_{nm}\\
A_{nm}^{(2)} = \sqrt{B \over 2} \left(X_1^{(2)} + i X_2^{(2)}\right)_{nm}
\end{array}
\end{equation}
commute:
\begin{equation}
\begin{array}{l}
\lb{aheis}
\left[A_{nm}^{(1)}, A_{n'm'}^{(1)\dag}\right]=\delta_{nm'}\delta_{n'm}\\
\left[A_{ij}^{(2)}, A_{i'j'}^{(2)\dag}\right]=\delta_{ij'}\delta_{i'j}.
\end{array}
\end{equation}
Inserting~(\ref{nepoly}) and~(\ref{acrel}) in~(\ref{yvacu}), 
we obtain
\begin{eqnarray}
\label{xvacu}
\lefteqn{
| K \rangle= 
\Big[ \epsilon ^{i_1...i_{M_1}} 
\left(\psi^{(1)\dag} \cos{\vap\ov 2} - 
\psi^{(2)\dag}\sin{\vap\ov 2}\right)_{i_1} 
\cdots }
\nonumber\\
\lefteqn{
~~~~~~~~\left\{ \left(\psi^{(1)\dag} \cos{\vap\ov 2} - 
\psi^{(2)\dag}\sin{\vap\ov 2}\right) 
\left(A^{(1)\dag}\cos{\vap\ov 2} 
- A^{(2)\dag} \sin{\vap\ov 2}\right)^{M_1-1}
\right\}_{i_{M_1}}\Big]^{K} }
\nonumber \\
\lefteqn{
~~~~~~~~~\Big[ \epsilon ^{j_1...j_{M_2}} 
\left(\psi^{(1)\dag} \sin{\vap\ov 2} + 
\psi^{(2)\dag}\cos{\vap\ov 2}\right)_{j_1} 
\cdots }
\nonumber\\
&
~~~~~~\left\{ \left(\psi^{(1)\dag} \sin{\vap\ov 2} + 
\psi^{(2)\dag}\cos{\vap\ov 2}\right) 
\left(A^{(1)\dag}\sin{\vap\ov 2} 
+ A^{(2)\dag} \cos{\vap\ov 2}\right)^{M_2-1}\right\}_{j_{M_2}}
\Big]^{K} | 0\rangle.
\end{eqnarray}

In what follows, we proceed without the use of the unitary 
transformation to construct the
wave function $ | \Phi \rangle $ describing the 
system of $M_1+M_2$ electrons at filling factor~(\ref{biff}). 
One has to realize a physical state  $ | \Phi \rangle $ 
that satisfies the Gauss law constraint~(\ref{glpo})
\begin{equation}
\label{state}
{\cal G} | \Phi \rangle =0   
\end{equation}
and allows us to establish a link with two well-known
wave functions. May be the best way to do this is 
to define two operators
\begin{equation}
\begin{array}{l}
A = A^{(1)} \otimes A^{(2)}\\
\psi= \psi^{(1)}\otimes \psi^{(2)}
\end{array}
\end{equation} 
where $\otimes$ is the tensor product.
Using these matrices of harmonic-oscillator 
operators, we build a vacuum configuration
\begin{eqnarray}
\label{nvacuum}
\lefteqn{
| \Psi \rangle = 
\Big[ \epsilon ^{i_1...i_{M_1}} \psi^{\dag(1)}_{i_1} 
\left(\psi^{(1)\dag}A^{(1)\dag}\right)_{i_2}\cdots
\left(\psi^{(1)\dag} A^{(1)\dag M_1-1}\right)_{i_{M_1}} 
\Big]^{K_{11}-K_{12}} }
\nonumber\\
\lefteqn{
~~~~~~~~\Big[ \epsilon ^{j_1...j_{M_2}} \psi^{(2)\dag}_{j_1} 
\left(\psi^{(2)\dag}A^{(2)\dag}\right)_{j_2}\cdots
\left(\psi^{(2)\dag} A^{(2)\dag M_2-1}\right)_{j_{M_2}} 
\Big]^{K_{22}-K_{12}} }
\nonumber\\
&
~~~~~\Big[ \epsilon ^{k_1...k_{M_1+M_2}} \psi^{\dag}_{k_1} 
\left(\psi^{\dag}A^{\dag}\right)_{k_2}\cdots
\left(\psi^{\dag} A^{\dag M_1+M_2-1}\right)_{k_{M_1+M_2}} 
\Big]^{K_{12}} | 0 \rangle.
\end{eqnarray}
which satisfies the Gauss law constraint~(\ref{glpo}) 
and therefore we have
\begin{equation}
\left(\psi^{(1)} \psi^{(1)\dag} + \psi^{(2)} \psi^{(2)\dag} - 
M_1K_{11} - M_2K_{22}\right) |\Phi \rangle =0.
\end{equation}
Novel about this vacuum configuration is that one can  
interpret the term 
\beq
\lb{INT}
\left[ \epsilon ^{k_1...k_{M_1+M_2}} \psi^{\dag}_{k_1} 
\left(\psi^{\dag}A^{\dag}\right)_{k_2}\cdots
\left(\psi^{\dag} A^{\dag M_1+M_2-1}\right)_{k_{M_1+M_2}} 
\right]^{K_{12}}
\eeq
as an inter-layer correlation. In conclusion, our
configuration could well be a good ansatz for 
the ground states of double--layered FQH fluids
in the formalism of the NCCS matrix model. This 
will be clarified in the next section.

\section{Link with literature}

Here we show how the Yoshioka--MacDonald--Girvin and
Halperin wave functions describing,
respectively, the double-layer
and the unpolarized QH systems 
can be recovered from our vacuum 
configuration~(\ref{nvacuum}).

Before starting, we note that for
any $N$-dimensional vector $\psi^{\dag}$ 
and $N\times N$ matrix $A^{\dag}$, the expression
of the form 
\beq
F(\psi^{\dag}, A^{\dag})=
\epsilon ^{i_1...i_{N}} \psi^{\dag(1)}_{i_1} 
\left(\psi^{(1)\dag}A^{(1)\dag}\right)_{i_2} \cdots
\left(\psi^{(1)\dag} A^{(1)\dag N-1}\right)_{i_{N}}
\eeq
has a one-to-one correspondence to the polynomial
\beq
f(z)= \epsilon ^{i_1...i_N}z_{i_1}^{0}\cdots z_{i_N}^{N-1}.
\eeq 
Now our task can be done by defining
a new complex variable
\begin{equation}
\label{zeta}
\zeta_i=\left\{ 
\begin{array}{l}
z_i^{(1)} \qquad {\mbox{for}}\; i=1, ... ,N\\
z_{i-N}^{(2)}\qquad {\mbox{for}}\; i=N+1, ... ,2N
\end{array}
\right.
\end{equation}
assuming that the particle numbers are 
equal, $M_1=M_2=N$, 
and recalling the Vandermonde determinant:
\begin{equation}
\prod_{i<j} \left(z_i-z_j\right) = {\rm det}\left(z_i^{N-j}\right)=
\epsilon^{i_1...i_N}z_{i_1}^{0}\cdots z_{i_N}^{N-1}.
\end{equation}
In terms of the complex coordinates,~(\ref{nvacuum})
reads
\begin{eqnarray}
\label{projection}
\lefteqn{
\Psi_{(K_{11},K_{22},K_{12})}=
\left[ \epsilon ^{i_1...i_N}  \left(z_{i_1}^{(1)}\right)^0 \cdots
\left(z_{i_N}^{(1)}\right)^{N-1} \right]^{K_{11}-K_{12}}~{}  }~~~~~~~~~~~~~~~~~~
\nonumber\\
\lefteqn{ 
~~\left[ \epsilon ^{j_1...j_N}  \left(z_{j_1}^{(2)}\right)^0 \cdots
\left(z_{j_N}^{(2)}\right)^{N-1} \right]^{K_{22}-K_{12}} }
\nonumber\\
&
\Big[ \epsilon ^{k_1...k_{2N}}  \zeta_{k_1}^0 \cdots
\zeta_{k_{2N}}^{2N-1} \Big]^{K_{12}}\; 
\Psi_{0}.\qquad 
\end{eqnarray}
It can be written in standard form as  
\begin{equation}
\lb{cvacu}
\Psi_{\left(K_{11},K_{22},K_{12}\right)}=
\prod_{i<j} \left(z_i^{(1)}-z_j^{(1)}\right)^{K_{11}}
\prod_{i<j} \left(z_i^{(2)}-z_j^{(2)}\right)^{K_{22}}
\prod_{i,j} \left(z_i^{(1)}-z_j^{(2)}\right)^{K_{12}} 
\; \Psi_{0}
\end{equation}
and now the inter-layer correlation
is
\beq
\prod_{i,j} \left(z_i^{(1)}-z_j^{(2)}\right)^{K_{12}}.
\eeq

Next, we will give two different applications 
of~(\ref{cvacu}).

\subsection{YMG wave functions}

Considering the two layers and treating them as
additional degrees of freedom, the $\nu={1\ov 2}$
state was predicted by Yoshioka, MacDonald and 
Girvin~\cite{girvin1}. They made
a straightforward generalization of the Laughlin
wave functions to those with the filling factor
\begin{equation}
\lb{ymgf}
\nu={2\over m+n}
\end{equation}
where $m$ and $n$ are integers. This can be obtained   
from our analysis by taking 
\begin{equation}
\begin{array}{l}
{K}=\left(\begin{array}{ll}
m&n\\
n&m\end{array}\right), \qquad
q=\left(\begin{array}{ll}
1&-1\end{array}\right)
\end{array}
\end{equation}
leading to the wave function 
\begin{equation}
\Psi_{(m,m,n)}=
\prod_{i<j} \left(z_i^{(1)}-z_j^{(1)}\right)^{m}
\prod_{i<j} \left(z_i^{(2)}-z_j^{(2)}\right)^{m}
\prod_{i,j} \left(z_i^{(1)}-z_j^{(2)}\right)^{n} \; \Psi_{0}.
\end{equation}
Choosing $m=3$ and $n=1$,
we recover the FQHE $\nu={1\over 2}$ state
corresponding to 
\begin{equation}
\Psi_{(3,3,1)}=
\prod_{i<j} \left(z_i^{(1)}-z_j^{(1)}\right)^{3}
\prod_{i<j} \left(z_i^{(2)}-z_j^{(2)}\right)^{3}
\prod_{i,j} \left(z_i^{(1)}-z_j^{(2)}\right)\; \Psi_{0}.
\end{equation}

\subsection{Halperin wave functions}

Another interesting result can be obtained. 
In the Halperin picture~\cite{halperin} in the 
context of single-layered unpolarized
QH systems, the labels $1$ and $2$
can be considered as an analogue of spin.
Following this idea, our bilayered system can
be seen as mixing layers of particles with  
spin up and spin down.

As a consequence, we obtain for $m=3$ and $n=2$ 
the unpolarized Halperin 
wave function with the filling factor
${2\ov 5}$ as
\begin{equation}
\Psi_{(3,3,2)}=
\prod_{i<j} \left(z_i^{(1)}-z_j^{(1)}\right)^{3}
\prod_{i<j} \left(z_i^{(2)}-z_j^{(2)}\right)^{3}
\prod_{i,j} \left(z_i^{(1)}-z_j^{(2)}\right)^{2}\; \Psi_{0}.
\end{equation}
This can be seen as a wave function of a system
of $N$ particles with spin parallel
and another $N$ particles with spin 
antiparallel to the external magnetic field.

\section{Conclusion}

We have developed a matrix model to describe 
bilayered QH systems at the filling 
factor $\nu=q_iK_{ij}^{-1}q_j$. 
The basic idea was to use two coupled harmonic-oscillators 
in a similar fashion as done by Susskind and Polychronakos.
Our model is a generalization
of their model and of course
reproduces its basic features by taking the coupling parameter 
$K_{12}$ to be zero.

Starting from an appropriate action we derived the equations
of motion for the different matrix model variables.
The corresponding Hamiltonian was obtained as the sum  
of free and interacting terms. A unitary transformation, 
more precisely a rotation around an angle $\varphi$,
led to a factorizing Hamiltonian.

Next,  we have constructed the ground states of the
system in two different ways. The first
was based on the unitary transformation
and from the ground state after rotation we 
have derived that before rotating the system.
The second was performed directly in terms of a combination 
of the matrices of harmonic-oscillator
operators of two layers. The obtained vacuum configuration
involved three different
quantities where one describes the 
inter-layer interaction. 

Subsequently, we have investigated the link between 
our second wave function and two others from literature.
After projecting the
vacuum configuration on the complex plane and 
using the Vandermonde determinant, we have shown 
how the Yoshioka--MacDonald--Girvin wave function
with the filling 
factor $\nu={2\over m+n}$ can be obtained 
from our model, in particular
that corresponding to the $\nu={1\over 2}$ state. 
Likewise, we have recovered the  
unpolarized Halperin wave function, 
especially that for the $\nu={2\over 5}$ state.

The case we have studied is in fact 
just a particular case of more 
general FQH states  where the fluid droplet is assumed to
consist of several coupled branches, say $M$ branches.
$M=1$ is the Laughlin (Susskind--Polychronakos) model,
$M=2$ is the model we have discussed here and 
$M\geq 3$ is the generic case, which can be seen
as a straightforward generalization of our case.

Of course still some important questions remain to be answered
e.g. about the fractional charge and statistics of
the particles and how to describe them in terms
of the proposed model. Another interesting question is related 
to the link between our model and 
Calogero and super--Calogero models. We will return to these issues
and related matter in future.

We close this section by noting that our model  
will be investigated in the forthcoming work~\cite{jellal4}
for the case of a single layer. Basically, we will
consider the Laughlin liquids in a confining potential that
is not of parabolic type and see how this affects the basic
features of the Susskind--Polychronakos model.

\section*{{Acknowledgments}}

We are very grateful to P. Bouwkgnet for
extremely helpful discussions. AJ's
work is supported by Deutsche Forschungsgemeinschaft
within the Schwerpunkt ``Quantum-Hall-Effekt''.

\end{document}